\documentclass{sig-alternate}

\usepackage{amsmath,amssymb,amsfonts}
\usepackage{algorithmic}
\usepackage{graphicx}
\usepackage{textcomp}
\usepackage{xcolor}
\usepackage{fancyhdr}

\usepackage[]{booktabs}
\usepackage[caption=false,font=footnotesize]{subfig}
\usepackage{siunitx}
\definecolor{codegreen}{rgb}{0,0.6,0}
\definecolor{codegray}{rgb}{0.5,0.5,0.5}
\definecolor{codepurple}{rgb}{0.58,0,0.82}
\definecolor{backcolour}{rgb}{0.95,0.95,0.92}
\usepackage{listings}

\usepackage{calc}
\graphicspath{{fig/}}

\newcommand{\smalltt}[1]{\texttt{\small #1}} 
\newcommand{\flushx}{\textsc{Flushx}} 

\usepackage{amssymb}

\usepackage{tikz}
\usetikzlibrary{tikzmark}

\usepackage{enumitem}
\setlist[enumerate]{label=\roman*}  

\usepackage{amsfonts}

\usepackage[T1]{fontenc}

\definecolor{Yellow}{rgb}{1,1, 0.6}
\definecolor{Red}{rgb}{1, 0.6, 0.6}

\usepackage{pgf}
\usepackage{colortbl}

\newcommand{\symA}{\raisebox{3\depth}{\small $\dag$}}
\newcommand{\symB}{\raisebox{3\depth}{\small $\ddag$}}

\usepackage{mathptmx} 

\usepackage{fancyhdr}
\usepackage[normalem]{ulem}
\usepackage[hyphens]{url}
\usepackage[sort,nocompress]{cite}
\usepackage[final]{microtype}
\usepackage[keeplastbox]{flushend}
\usepackage[bookmarks=true,breaklinks=true,letterpaper=true,colorlinks,linkcolor=black,citecolor=blue,urlcolor=black]{hyperref}

\usepackage{cleveref}
\crefformat{section}{\S#2#1#3} 
\crefformat{subsection}{\S#2#1#3}
\crefformat{subsubsection}{\S#2#1#3}

\fancypagestyle{firstpage}{
  \fancyhf{}
  
  \fancyfoot[C]{\thepage}
}

\title{SIMF: Single-Instruction Multiple-Flush Mechanism for Processor Temporal Isolation} 

\author{
Tuo Li$\symA$, Bradley Hopkins$\symB$, and Sri Parameswaran$\symA$\\
       \affaddr{$\symA$School of Computer Science and Engineering, University of New South Wales, Australia}\\
       \email{\{tuoli,sri.parameswaran\}@unsw.edu.au}\\
       \affaddr{$\symB$Defence Science and Technology Group,
Edinburgh, Australia}\\
       \email{\{brad.hopkins\}@dst.defence.gov.au}
}

\begin{document}
\maketitle
\thispagestyle{firstpage}
\pagestyle{plain}


\begin{abstract}

Microarchitectural timing attacks are a type of information leakage attack, which exploit the time-shared microarchitectural components, such as caches, translation look-aside buffers (TLBs), branch prediction unit (BPU), and speculative execution, in modern processors to leak critical information from a victim process or thread.
To mitigate such attacks, the mechanism for flushing the on-core state is extensively used by operating-system-level solutions, since core-level state is too expensive to partition. In these systems, the flushing operations are implemented in software (using cache maintenance instructions), which severely limit the efficiency of timing attack protection.

To bridge this gap, we propose specialized hardware support, a single-instruction multiple-flush (SIMF) mechanism to flush the core-level state, which consists of L1 caches, BPU, TLBs and register file. We demonstrate SIMF by implementing it as an ISA extension, i.e., \flushx{} instruction, in scalar in-order RISC-V processor. The resultant processor is prototyped on Xilinx ZCU102 FPGA and validated with state-of-art seL4 microkernel, Linux kernel in multi-core scenarios, and a cache timing attack. Our evaluation shows that SIMF significantly alleviates the overhead of flushing by more than a factor of two in execution time and reduces dynamic instruction count by orders-of-magnitude.

\end{abstract}

\section{Introduction}

\begin{figure}[tb]
\centering
  \subfloat[Process switch (Alice $\to$ Bob) in a processor core]{
  \includegraphics[width=.6\columnwidth]{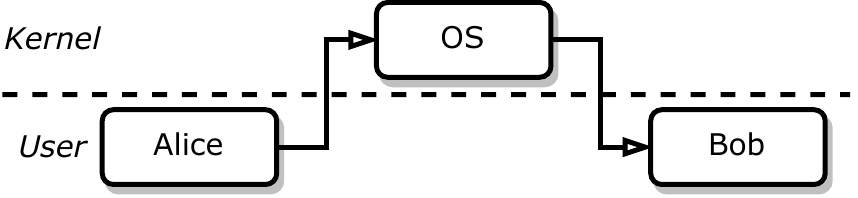}\label{fig:switch}
  }\\
  \subfloat[Leakage from persistent state at core level. Grey color denotes residual state of the victim Alice; $t_i$: time slice $i$.]{
  \includegraphics[width=.75\columnwidth]{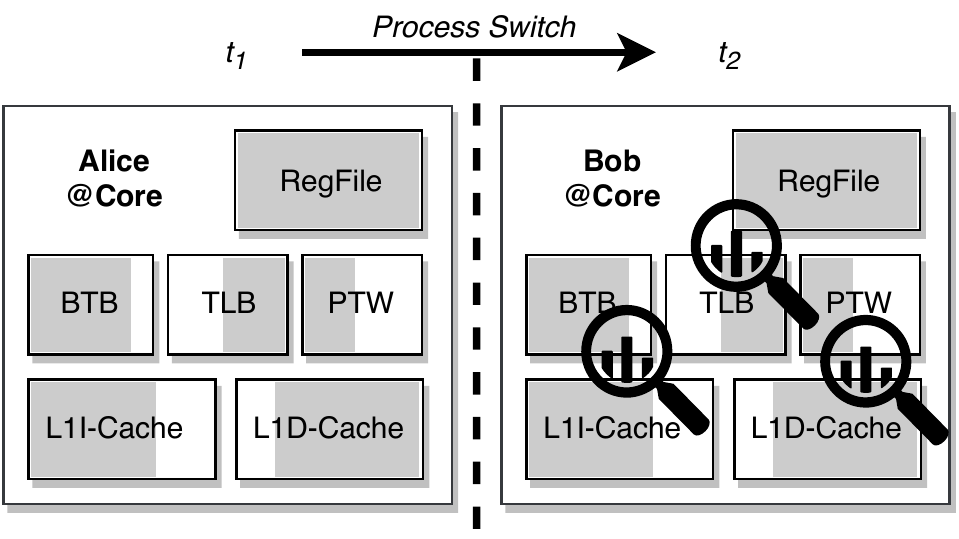}\label{fig:state}
  }
  \caption{Information leakage in time-shared hardware.} 
\label{fig:problem}
\vspace{-10pt}
\end{figure}

One of the fundamental techniques to secure computer program execution is the confinement of programs to eliminate or minimize  information leakage~\cite{Lampson:1973:NCP}. 
To improve computing performance, contemporary computer systems contain increasingly complex architectures and microarchitectures (shown in Figure~\ref{fig:problem}), which allows for the time-sharing of hardware (much of the hardware is shared with multiple processes), faster memory references, reduced branch penalty etc. These architectural improvements provide unique opportunities for an adversary to orchestrate advanced information leakage attacks upon modern processing systems~\cite{Cock:2014:LME,Hennessy:2019:NGA}.

Microarchitectural timing attack (MTA) is an information leakage attack which has been recently shown to be successful~\cite{ Canella2019}. Example exploits include, Meltdown~\cite{Lipp2018meltdown} and Spectre~\cite{KGH2018}, which were mounted by taking advantage of branch prediction, speculative execution, and caches.

In such an attack, the adversary typically relies on the temporal interference of the persistent state in certain microarchitectural components (such as cache) of processors, to create timing variations (e.g., cache access time difference). Such timing variations, along with a priori knowledge of the victim system (software and hardware), can be used to infer the secret data in a time-shared system's hardware~\cite{Yarom:2016}.

In order to mitigate MTA, existing hardware-based techniques focus on redesigning the cache (which targets cache timing attack) as well as adding information flow tracking across a processor by drastically modifying  the processor architecture, which must be created with a corresponding new hardware description language.  Software-based techniques at hypervisor or OS level seek to maximally close the cache timing channels by cache partitioning (at shared caches), cache flushing (at private caches). To mitigate MTA system-wide (including all the possible timing channels in  multiple microarchitectural states), software-based methods, such as \emph{temporal isolation} (or, time protection)~\cite{Ge:2019:TPM} propose the  flushing of all on-core states, including private caches, translation look-aside buffers, branch prediction unit, etc. 

Recent software-based methods~\cite{Zhang:2013:DRC,Ge:2019:TPM,Oleksenko:2018:VPS}  heavily rely on the flushing mechanism. While \emph{flushing or cleansing} the persistent state of multiple vulnerable hardware components at the core level (within Level 1) is regarded as indispensable for creating sound and complete temporal isolation system-wide, these studies have also shown the incompetency of the flushing instructions provided by the existing ISAs, in particular when flushing must be performed frequently (up to 50 KHz). 
The limitations of the \textbf{existing hardware support} for flushing are discussed in-depth in \cref{sec_bg_isa} and summarized  as the follows:
1) as shown in Figure~\ref{fig:arm_code} numerous lines of instructions are used to create routines or functions for flushing each and every microarchitectural component of interest in sequence  (sequence is enforced by inserting explicit barrier instructions in between the flushing instructions); and,
2) individual instructions used for flushing may not contain all necessary components, for example, x86 provides one instruction to flush the entire cache hierarchy but lacks a dedicated L1 cache flushing instruction or an instruction to flush the branch prediction unit. 
As elaborated in \cref{sec_motive} and \cref{sec_bg} (Figure~\ref{fig:tickvscost}), in our initial estimation, flushing at high frequency can incur more than 30\% of additional dynamic instructions, which is equivalent to significant execution time and power overhead (instruction fetch power takes approximately ~30\% of CPU power~\cite{Bose2003}).

In this paper, we aim to create one individual specialized flushing instruction to raise the efficiency of temporal isolation at the core level for on-core state towards minimizing the existing and potential timing channels~\cite{Ge2018}. Based on this instruction, we also aim to explore the upper-bounds of the set of software-based MTA mitigations, which are constructed on top of the flushing mechanism, in terms of efficiency. 

To this end, we first propose a single-instruction multiple-flushing (SIMF) scheme, which integrates flushing operations in a single instruction to clear the  core-level state. The key advantage of SIMF is: 1) sharply reducing the dynamic instruction count (leading to cycle counts and instruction fetch power) dedicated to flushing; 2) minimal extension (adding one instruction to ISA) to the existing hardware; 3) implicitly enforcing the orders of flushing operations in one instruction, without using explicit barrier instructions; 4) programming benefits, including strong atomicity, which cannot be interrupted in the middle (for the case discussed in~\cite{Zhang:2013:DRC}) and simplicity. 

For thoroughly investigating SIMF, we prototyped SIMF in an open-source scalar in-order RISC-V\footnote{\url{riscv.org}} processor. We extended RISC-V ISA with one additional instruction called \flushx{}, which flushes the core-level state including L1/L2 TLBs, L1 caches, branch prediction unit (BTB, RAS, BHT). We also explored flushing the register file, which has been reported as a potential information leakage target~\cite{Canella2019}. Our evaluation shows that SIMF significantly alleviates the overhead of flushing by more than a factor of two in execution time and reduces dynamic instruction count by orders-of-magnitude.

The main contributions of this paper are as follows:
\begin{itemize} 
\item An instruction extension, which can flush L1 caches, TLBs, BPU, and register file in one instruction execution;
\item A FPGA prototype of SIMF in multi-core RISC-V processor;
\item Integration of SIMF in state-of-art seL4 microkernel and Linux kernel;
\item An evaluation with seL4 kernel, Linux kernel (multi-core), and \smalltt{Prime+Probe} cache timing attack, on FPGA.  
\end{itemize}

\section{Motivational Example}\label{sec_motive}
\lstset{
basicstyle=\scriptsize\ttfamily,
columns=flexible,
breaklines=true,
numberstyle=\tiny\ttfamily,
numbers=left,
stepnumber=1,
numbersep=5pt,
framexleftmargin=1.5em,
xleftmargin=1.5em,
escapechar=@,
language=c
}

\begin{figure}[!t]
\centering
\begin{lstlisting}
void cleanInvalidateL1Caches(void){//flush L1 caches
    dsb();@\label{line:calldsb1}@
    cleanInvalidate_D_PoC();@\label{line:callfl1dc}@
    dsb();@\label{line:calldsb2}@
    invalidate_I_PoU();@\label{line:callfl1ic}@
    dsb();@\label{line:calldsb3}@
    }
void cleanInvalidate_D_PoC(void){//flush L1 D-cache@\label{line:deffl1dc}@
    int clid = readCLID();
    int loc = LOC(clid);
    int l;
    for (l = 0; l < loc; l++) {
        if (CTYPE(clid, l) > ARMCacheI) {
            word_t s = readCacheSize(l, 0);
            int lbits = LINEBITS(s);
            int assoc = ASSOC(s);
            int assoc_bits = wordBits - clzl(assoc - 1);
            int nsets = NSETS(s);
            int w;
            for (w = 0; w < assoc; w++) {
                int s;
                for (s = 0; s < nsets; s++) {
                    cleanInvalidateByWSL((w << (32 - assoc_bits)) | (s << lbits) | (l << 1));
                }}}}}
static inline void invalidate_I_PoU(void){//flush L1 I-cache@\label{line:deffl1ic}@
    asm volatile("ic iallu");
    isb();
    }
\end{lstlisting}
\caption[]{Source code of the key functions for ARMv8 processor L1 cache flush in a contemporary operating system microkernel~\cite{Klein_AEMSKH_14} (\url{https://sel4.systems/}).}\label{fig:arm_code}
\end{figure}

\subsection{seL4's L1 Flush with ARM ISA} 
Figure~\ref{fig:arm_code} presents the code snippet of software functions, which perform flushing L1 caches for temporal isolation in the seL4 microkernel, which targets the ARMv8 architecture. In seL4, \texttt{\small cleanInvalidateL1Caches} (Line 1-8) is called for flushing L1 caches. This function is composed of two major steps, which are essentially two function calls:
one for flushing L1 data cache (\texttt{\small cleanInvalidate\_D\_PoC} at Line~\ref{line:callfl1dc}),
and one for flushing L1 instruction cache (\texttt{\small invalidate\_I\_PoU} at Line~\ref{line:callfl1ic}).
In addition, three data synchronization barriers are called (\texttt{\small dsb} at Line~\ref{line:calldsb1}, \ref{line:calldsb2}, \ref{line:calldsb3}) before and after the function calls for flushing L1 D- and I-cache.

\begin{figure}[!t]
\centering
\includegraphics[width=\columnwidth]{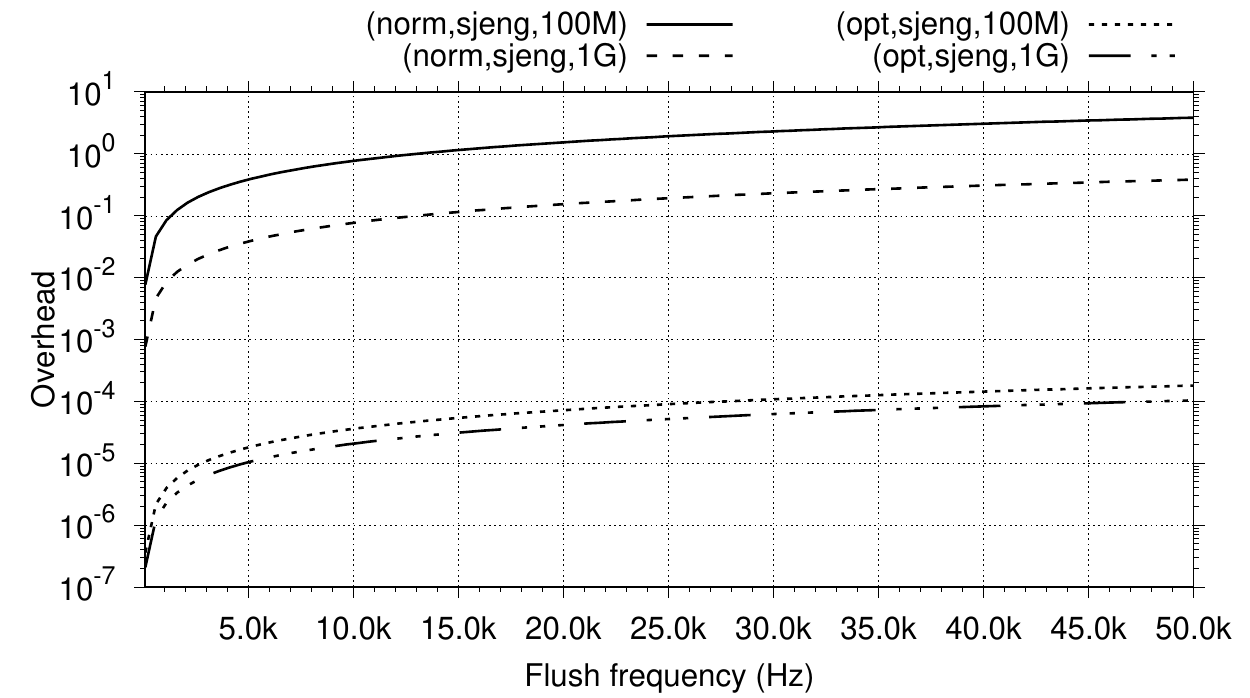}
\caption{Estimated instruction count overhead of flushing mechanism in Figure~\ref{fig:arm_code} as a function of flushing frequency (100 Hz to 50 KHz) in one SPECint program, with two clock frequencies (100 MHz and 1 GHz).} 
\label{fig:tickvscost}
\end{figure}

Figure~\ref{fig:tickvscost} further depicts the dynamic instruction overhead resulted by seL4's flushing mechanism ``\smalltt{cleanInvalidateL1Caches}'' in Figure~\ref{fig:arm_code}. Since we only consider L1 caches in this example, the resultant overhead is lower than the actual overhead of core-level flushing.
The overhead (y-axis) is shown as a function of flushing frequency (x-axis). 
First, we estimate the dynamic instruction count of executing \texttt{\small cleanInvalidateL1Caches} on ARMv8, by analyzing the relevant code segments (referencing the functions in Figure~\ref{fig:arm_code}) in objdump output of the seL4. 
The size of the main loop in \texttt{\small cleanInvalidate\_D\_PoC} depends on the cache configuration of L1 D-cache in the system --- 
the number of sets \smalltt{nsets} and the number of ways \smalltt{assoc}.
Here, we assume the L1 D-cache has the following configuration: $\smalltt{nsets} = 64$ and $\smalltt{assoc} = 8$.

Based on the above assumption, the total count of dynamic instructions for flushing operation in a program is approximated in the following way:
1) the dynamic instruction count and cycle count of normal program execution are obtained via FPGA emulation; 
2) operating frequency, flushing frequency, and cycle count are used to calculate the number of flushes performed during the program execution; and, 
3) the total dynamic instruction count of flushes is normalized to the total dynamic instruction count of normal program execution.

To briefly envision the scaling of the overhead,  
we consider two representative operating clock frequencies, 100 MHz and 1 GHz, 
as well as one SPECint2006 workload, 458.sjeng.
The flushing frequency is defined as $x \in [100,50000]$ Hz, considering the values that are potential and is adopted in practice~\cite{Ge:2019:TPM,Zhang:2013:DRC}.
To contrast with the existing flushing mechanism, we also estimate the overhead with the optimal flushing mechanism in theory, which realizes the complete flushing operations in one instruction, for 458.sjeng. The optimal mechanism is denoted as ``opt'' (whereas ``norm'' denotes normal system) for the first element in the tuple in the legend of Figure~\ref{fig:tickvscost}.

\subsection{Observation}
Figure~\ref{fig:tickvscost} provides two principal observations:
First, the dynamic instruction count overhead can become substantial as flushing frequency increases. In the case of 100 MHz clock frequency, core-level flushing incurs about 10\% overhead with 1 KHz flushing frequency and almost 100\% overhead when approaching 10 KHz. For 1 GHz clock frequency, the same flushing frequency results in much fewer occurrences of flushing. Hence, the overhead is less prominent, however still reaches 10\% at 10 KHz and approaching 100\% at 50 KHz.     
Second, the optimal design of flushing mechanism can lead to a sharp decrease of orders-of-magnitude in dynamic instruction count overhead. Such a huge gap between ``opt'' and ``norm'' suggests significant promise in performance and power efficiency. 
Flushing operations are performed during context switches. Context switches usually lead to refilling the on-core state (e.g., caches), therefore, the cold state, caused by the flushing during context switches, incurs minimal additional performance overhead~\cite{Ge:2019:TPM}.

The aforementioned observations and discussion strongly motivate the design of a special compound instruction to efficiently flush the core-level state residing in multiple components in processor architecture.

\section{Background}\label{sec_bg}
\subsection{Threat Model}
The threat model in this study is defined below and is similar to the ones defined in existing advanced microarchitectural timing attacks~\cite{KGH2018} and research~\cite{Ge:2019:TPM}:
1) the adversary Bob is assumed to have control of a user process, which shares the same processor core with the victim Alice's process, in a contemporary preemptive operating system;
2) based on time-sharing mechanisms (e.g., preemption), Bob aims to manipulate the core-level state, such as L1 caches, to exploit the temporal dependence of the state (e.g., cache contention) between Bob and Alice's time slice, hence rendering the timing channels;
3) Bob is capable of designing the point of time to invoke the preemption or inter-process interrupt (IPI), such as the attack discussed in~\cite{Neve2007}, to carefully target a specific part of Alice's execution; and, 
4) the system is equipped with necessary temporal isolation mechanisms for protecting L2 caches and LLC, such as~\cite{Ge:2019:TPM}.

\subsection{Temporal Isolation}

\begin{figure}[tb]
\centering
\includegraphics[width=\columnwidth]{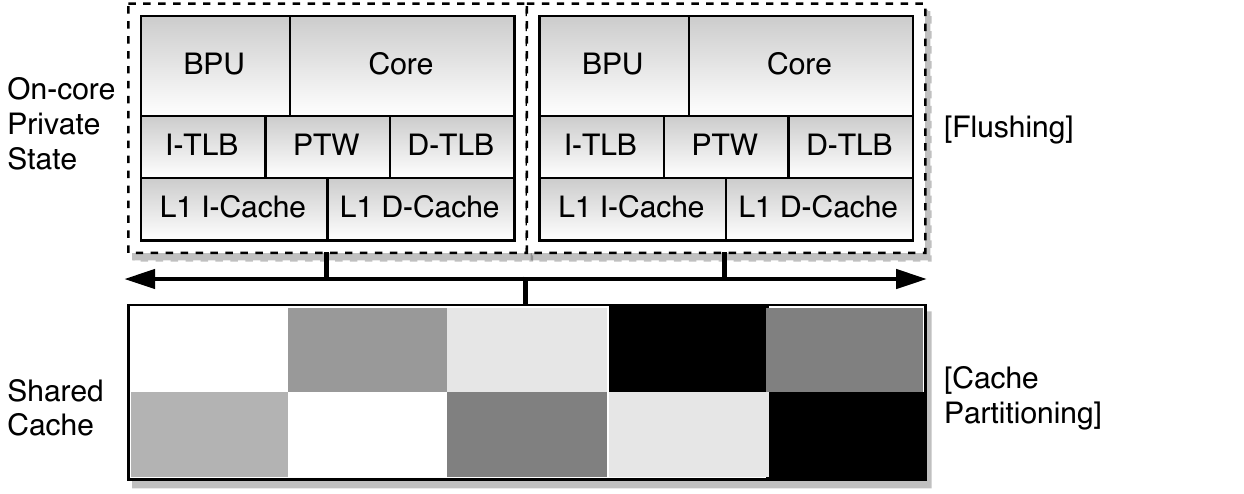}
\caption{MTA mitigation (temporal isolation) at OS level.} \label{fig:temporaliso}
\end{figure}

As shown in Figure~\ref{fig:temporaliso}, temporal isolation~\cite{Bui:2011:TIM} prevents temporal interference, hence eliminates timing channels, by spatially or temporally partitioning the hardware resources.
Existing OS-level temporal isolation methods generally use cache partitioning (page coloring) for shared large caches, such as L2 and last-level cache (LLC), which typically are indexed by the physical address. This way is also referred as spatial partitioning.

However, applying cache partitioning to core-level ``private'' resources, e.g., L1 caches and TLB, which are indexed by virtual address is difficult. For enforcing temporal isolation in these components, flushing is advocated to cleanse the state in these components~\cite{Ge:2019:TPM}. By cleansing the state, there will be no residual state left by the victim for the attacker to create timing channels. This way is also called temporal partitioning.

The frequency of core-level state flushing is determined by the temporal isolation scheme. The seL4's time protection~\cite{Ge:2019:TPM} initiates flushing upon every security-domain switch, which is approximately once per 10-100 milliseconds. 
seL4's domain is a group of threads, and hence, less frequent than the thread switch. The attack from the threads in the same domain is presumed to not happen. 
In other schemes, such as D\"uppel~\cite{Zhang:2013:DRC}, flushing is performed as noise injection to prevent cache timing channel. Therefore, flushing is activated ``at least 50 times per millisecond''. 

\begin{table}[!t]\scriptsize
\caption{Support for core-level flushing in existing ISAs (light gray: overkill; dark gray: underkill)}
\begin{center}
\begin{tabular}{c|cccc}
\toprule%
\textbf{ISA}	& \textbf{L1DC} & \textbf{L1IC} & \textbf{TLB} & \textbf{BP}\\
\midrule%
\textbf{x86}	& \cellcolor[gray]{0.8}\texttt{wbinvd} & \cellcolor[gray]{0.8}\texttt{wbinvd} & \texttt{invpcid} &{\color{white}\cellcolor[gray]{0.2} indirect}\\
\textbf{ARM32}	& \texttt{dccisw} & \texttt{iciallu} & \texttt{tlbiall} & \texttt{bpiall} \\
\textbf{ARM64}	& \texttt{dc cisw} & \texttt{ic iallu} & \texttt{tlbi allex} & {\color{white}\cellcolor[gray]{0.2} n/a }\\
\bottomrule%
\end{tabular}
\end{center}
\label{tab_isa_support}
\end{table}%

\subsection{Existing ISA Support for Flushing}\label{sec_bg_isa}

In existing temporal-isolation systems~\cite{Ge:2019:TPM}, the implementation of the flushing mechanism is based on existing instructions, such as cache maintenance instructions, provided by the ISA.
Table~\ref{tab_isa_support} shows how the flushing operations at core level can be implemented with ARM and x86 architectures.
\begin{figure*}[tb]
\centering
\includegraphics[width=0.9\textwidth]{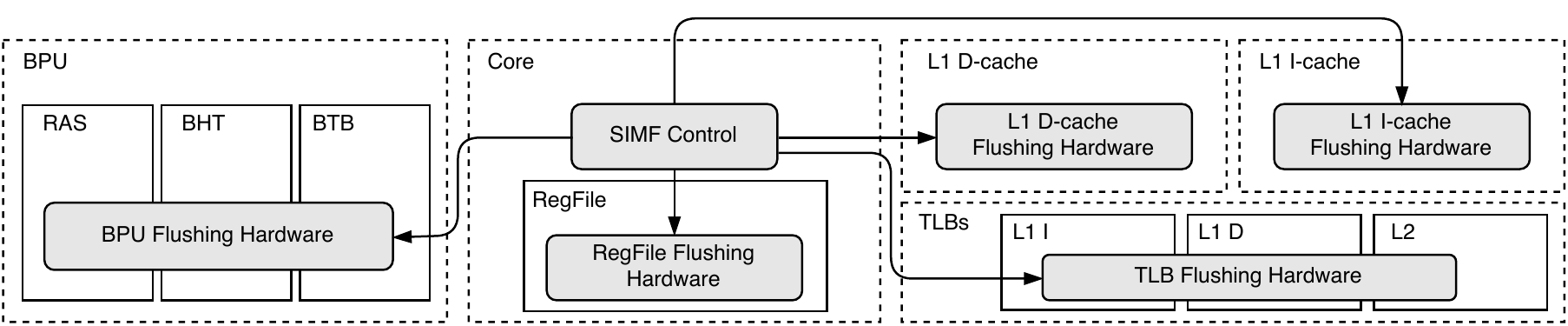}
\caption{Single-instruction multiple-flush mechanism overview.} \label{fig:simfov}
\end{figure*}

\textbf{x86 ISA support for flushing:}
\smalltt{clflush} invalidates and writes back (if dirty) the cacheline that contains the virtual address specified with the source operand for entire cache hierarchy (data and instruction). \smalltt{clflush} is not privileged, and can be executed in user mode.
\smalltt{wbinvd} invalidates and writes back (if dirty) cachelines for entire cache hierarchy. \smalltt{wbinvd} is privileged, and can only be executed in system mode.
\smalltt{invlpg} invalidates any TLB entries specified with the source operand. 
\smalltt{invlpg} is privileged, and cannot be deployed in user programs.
\smalltt{invpcid} invalidates mappings in the TLBs and paging-structure caches based on register operand.
x86 has no dedicated instruction to flush the state in branch prediction unit (BPU).
To achieve BPU cleansing, the indirect branch control feature~\cite{Intel2018} must be adopted.

\textbf{ARM ISA support for flushing:}
\smalltt{dc\textvisiblespace cisw} invalidates and writes back the specified individual data cacheline (in set/way format, instead of virtual address format) at the specified cache level for ARM64 (aarch64).
\texttt{\small dccisw} is one of ARM's cache maintenance instructions.
This instruction is privileged and hence must be executed in system mode. The equivalent instruction for ARM32 (aarch32) is \texttt{\small dccisw}.
\smalltt{ic\textvisiblespace iallu} invalidates entire L1 I-cache (to the point of unification, i.e., L2) for ARM64 (aarch64). This privileged instruction also invalidates BPU. The equivalent instruction for ARM32 is \texttt{\small iciallu}.
\smalltt{tlbi\textvisiblespace alle$x$} ($x \in \{1,2,3\}$) invalidates all entries in TLB for the specified translate regime (i.e., EL1, EL2, EL3). This set of instructions are ARM64 system instructions, while the ARM32 equivalent is one instruction, \texttt{\small tlbiall}.
\smalltt{bpiall} is an ARM32 system instruction, which invalidates all the entries in BPU. The ARM64 equivalent does not exist.

\textbf{RISC-V ISA support for flushing:}
As an emerging ISA, RISC-V ISA is rapidly including new instructions. There are two privileged instructions for flushing:\footnote{Rocket chip recently added an optional L1 d-cache flushing instruction, \texttt{CFLUSH.D.L1}, which is "only for power-down'' and ``... only supported on systems without S-mode.'' This instruction is not available in the version of rocket chip used in this paper.}
1) \smalltt{fence.i} for synchronizing the instruction and data streams, which essentially flushes the L1 I-cache;
2) \smalltt{sfence.vma} for flushing the TLBs.

\section{SIMF Mechanism}

As shown in Figure~\ref{fig:simfov}, 
SIMF mechanism design includes: 1) hardware of flushing operations for each on-core state; 2) the control logic, which manages the issue order of the flushing operations; and, 3) the instruction deployment in software.

In this paper, for the sake of brevity, we consider an in-order scalar processor architecture as the base system, which does not include prefetcher hardware. However, the implementation would not be too onerous for an out of-order processor. The principles remain similar.

\subsection{Sphere of Flushing}

Taking into account the study in ~\cite{Ge2018}, about microarchitectural timing channels, SIMF targets a comprehensive set of states on-core, which we define as the sphere of flushing (SoF). 
SoF includes all the on-core state, L1 D-Cache, L1 I-Cache, TLB and the BPU.

\textbf{L1 D-Cache} is prone to timing channel~\cite{Hu:1992:LSC}. Cache access time can vary substantially depending on whether the access is a hit or miss. An attacker can exploit the access time variation to mount timing channel attacks, for example, as shown using \textsc{Prime+Probe}~\cite{Osvik:2006:CAC,Percival05cachemissing}.

\textbf{L1 I-Cache} is similar to L1 D-cache, in terms of cache time access variation. The main difference is that L1 I-cache timing attack relies on creating cache contention via carefully crafted control flow and seeks to leak the execution flow of the victim process (e.g., cryptographic algorithms)~\cite{Aciicmez:2007:YMA}. 
 
\textbf{TLB} timing channel can be constructed by exploiting the time variation of a virtual address translation between TLB hit and miss~\cite{Hund:2013:PTS,Gras2018} for control flow (instruction TLB) and data flow (data TLB). We assume a contemporary two-level TLB microarchitecture ( e.g., Intel Nehalem\footnote{\url{www.intel.com/pressroom/archive/reference/whitepaper_Nehalem.pdf}} and RISC-V Rocket~\cite{Asanovic:EECS-2016-17}).

\textbf{BPU} incurs varied time cost between a correct and wrong prediction.
An attacker can mount the timing channel attack based on this time variation to obtain the execution flow of the victim process~\cite{Aciicmez2007}.

\textbf{RegFile} is not targeted by existing OS-level techniques~\cite{Ge:2019:TPM}. It has been discussed as a source of exploitable time variation, when there are data dependencies between registers~\cite{Coppens2009}, as well as can leak the data stored in the registers~\cite{Stecklina2018}.

\begin{figure}[tb]
\centering
\subfloat[L1 D-Cache]{
	\includegraphics[width=0.7\columnwidth]{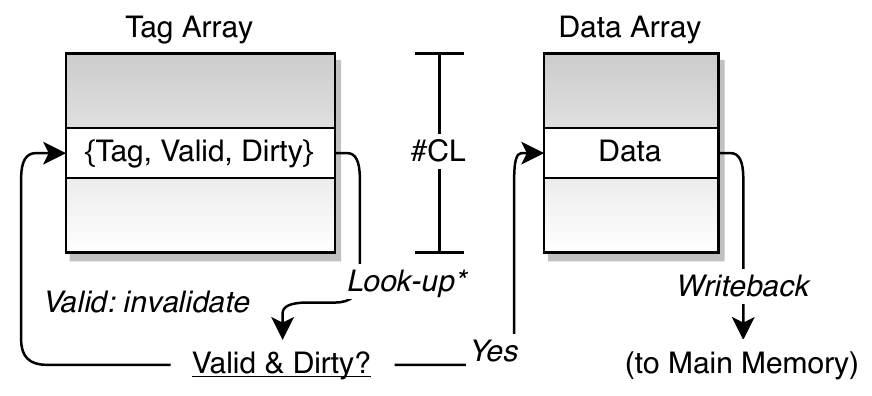}\label{fig:l1dcflushop}
}
\subfloat[L1 I-Cache]{
\includegraphics[width=0.25\columnwidth]{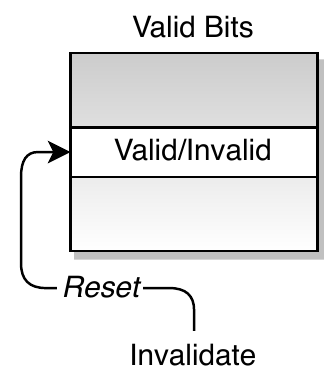}\label{fig:l1icflushop}
}
\\
\subfloat[TLBs]{
\includegraphics[width=0.98\columnwidth]{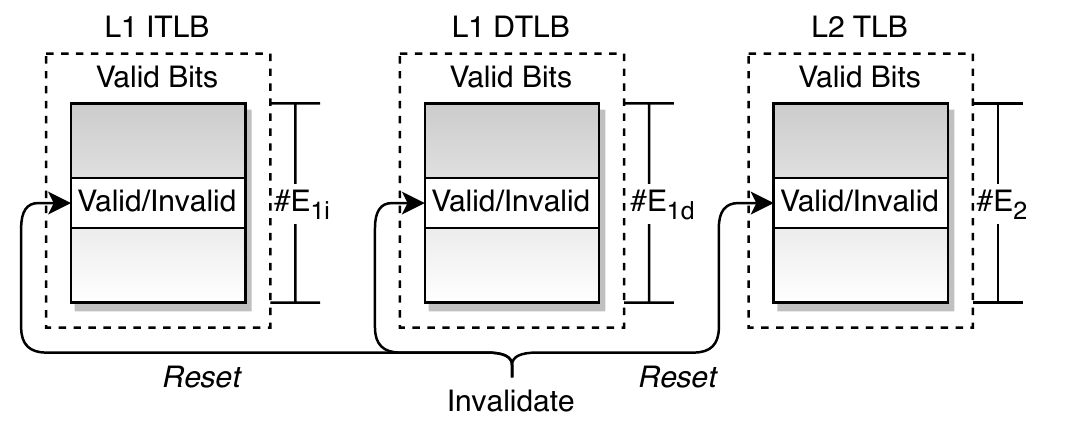}
\label{fig:tlbflushop}
}
\\
\subfloat[BPU]{
\includegraphics[width=0.9\columnwidth]{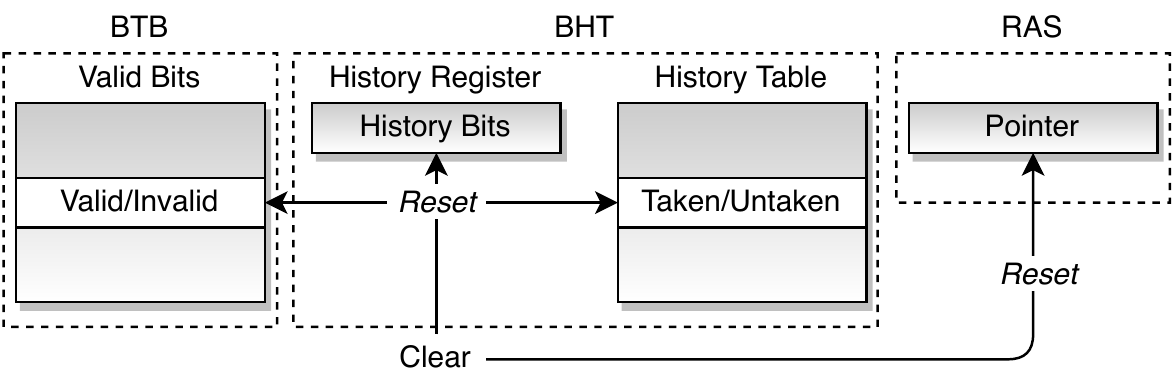}
\label{fig:bpuflushop}
}
\caption{SIMF flushing operations.} \label{fig:flushop}
\end{figure}

\subsection{Flushing Operations}\label{sec_fo}

Figure~\ref{fig:flushop} depicts SIMF's flushing operations targeting the components in SoF. Depending on the microarchitecture of each component, the flushing operations vary.  

\textbf{Flushing L1 D-cache} (Figure~\ref{fig:l1dcflushop}) mainly aims to flush the valid and dirty bits.
These bits represent validity and coherency status of each cache line.
Due to data coherence, clearing each cache line's valid bits and dirty bits must be performed along with write-back operation, if the cache line is dirty.
Therefore, the entire L1 D-cache flush includes three sub-operations:
1) look up the cache line status corresponding to one selected cache line from the tag array;
2) write back the cache line, if it is dirty; and,
3) resetting the valid and dirty bits.
These steps must be done for every cache line in the cache,
which totals to \smalltt{\#CL}. Hence, L1 D-cache flushing comprises $N$ ($N = \smalltt{\#CL}$) sequences of the sub-operations mentioned above.

\textbf{Flushing L1 I-cache} (Figure~\ref{fig:l1icflushop}) is a simple form of flushing L1 D-cache, since it is not necessary for I-cache to maintain coherence. Flushing L1 I-cache operation only needs to invalidate or reset the valid bits, where each bit maps to one cache line.

\textbf{Flushing TLBs} (Figure~\ref{fig:tlbflushop}) includes flushing TLBs at Level 1 and Level 2. 
This TLB system includes separate L1 TLBs for instruction (L1 ITLB) and data (L1 DTLB) and one unified L2 TLB (L2 TLB).
TLB flushing operation aims to clean the valid bits for each TLB entry.

We assume BPU uses two-level adaptive branch prediction~\cite{Yeh:1991:TAT}. The operation for \textbf{flushing BPU} (Figure~\ref{fig:bpuflushop}) is composed of three major sub-operations as follows:
1) flushing branch target buffer (BTB) aims to clear the valid bits in BTB;
2) flushing branch history target (BHT) is more complex than BTB, where the history bits in the history register and the entries in the history table are to be cleared; and,
3) flushing return address stack (RAS) can be realized by resetting the stack pointer, which indicates that the RAS is empty.

\textbf{Flushing RegFile} aims to remove the user data directly from RegFile. Given a RegFile consisting of \smalltt{\#Regs} registers, \smalltt{\#Regs} registers are cleared.
In addition, for maintaining the program's correctness, RegFile flushing must be carefully operated after the registers are saved in cache/memory. Vice versa, the registers need to be recovered after flushing, before continuing the program execution.
Hence, the context switch (for process switch) is a convenient point of time, when RegFile flushing can be performed. 

\subsection{Merging Flushing Operations}\label{sec_simfdef}

Based on the design of the flushing operations with respect to the SoF components, SIMF control is realized by merging these operations into one instruction.

Operation merging or fusion problem can be briefly defined as follows:
Given the set of predefined flushing operations (FOs), noted as $\mathcal{O} = \{O_1, O_2, ..., O_N\}$ and pipeline stages $\mathcal{S} = \{S_{1}, S_{2}, S_{3}, ..., S_{M}\}$, scheduling $O_i$ into pipeline stage $S_j$.

\begin{figure}[tb]
\centering
\includegraphics[width=0.8\columnwidth]{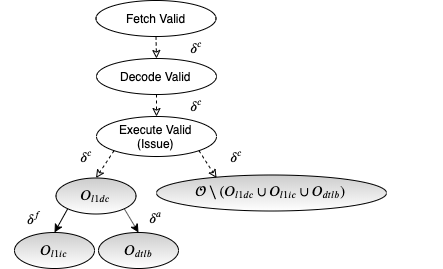}
\caption{Dependency graph for flushing operations in SIMF (``$\setminus$'': set minus).} \label{fig:cdfg}
\end{figure}

Figure~\ref{fig:cdfg} summarizes the design constraints, represented as control and data dependencies, which must be considered when merging flushing operations in one instruction.
The control dependencies are denoted as $\delta^c$. The data dependencies are denoted as $\delta^f$ for flow/true dependency, also known as read-after-write (RAW), as well as $\delta^a$ for anti-dependency, a.k.a. write-after-read (WAR). SIMF considers two main data dependencies: 1) Flushing L1 I-cache ($O_{l1ic}$) will cause refilling the instructions, whose newest version might be in the L1 D-cache and hence depends on flushing L1 D-cache ($O_{l1dc}$); and, 2) Flushing L1 D-cache might require address translation in D-TLB, which will be flushed by $O_{dtlb}$.  

Table~\ref{tab:fusion} depicts how the proposed SIMF merges the flushing operations
$\mathcal{O}$, assuming a classic 5-stage pipeline $\mathcal{S} = \{S_{IF}, S_{ID}, S_{EX}, S_{ME}, S_{WB}\}$.
Given the constraints, we aim to schedule the FOs as late as possible (ALAP), which is equivalent to prioritizing issuing the FOs near the commit (i.e., write-back) stage.
The first three stages are not considered given the control constraints (flushing must not start before valid execute/issue stage).
$O_{l1dc}$ is scheduled one stage earlier than the other operations ($\mathcal{O}\setminus O_{l1dc}$) to ensure that the dependencies are satisfied.



\begin{table}[!t]\scriptsize
\caption{SIMF scheduling in a classic 5-stage pipeline based on dependencies in Figure~\ref{fig:cdfg} (CCs: clock cycles, $\alpha$: cycles per cacheline)}
\begin{center}
\begin{tabular}{c | cc}
$\mathcal{S}$	& \textbf{FOs} & \textbf{\#CCs} \\
\midrule%
\textbf{IF}	& 
$\varnothing$ & 
	\texttt{n/a} \\
\textbf{ID}	& 
$\varnothing$ & 
	\texttt{n/a} \\
\textbf{EX}	& 
$\varnothing$ & 
	\texttt{n/a} \\
\textbf{ME}	& $O_{l1dc}$ & $\sim\alpha\cdot\#CL$ \\
\textbf{WB}	& $\mathcal{O} \setminus O_{l1dc}$ & 1 \\
\end{tabular}
\end{center}
\label{tab:fusion}
\end{table}%

\subsection{SIMF Pipeline Control}\label{sec_pctrl}

\begin{table}[!t]\scriptsize
\caption{SIMF Pipeline Control (SIMF instruction highlighted in gray)}
\centering
\setlength\tabcolsep{4pt} 
\begin{tabular}{c|ccccccccc}
\toprule%
\textbf{I} $ \backslash$\textbf{T} & $\textbf{CC}_{0}$&$\textbf{CC}_1$ &
	$\textbf{CC}_2$ & $\textbf{CC}_3$ & $\textbf{CC}_4$ & $\textbf{CC}_5$
	& $\textbf{CC}_6$ & $\textbf{CC}_7$ & $\textbf{CC}_8$\\
\midrule%
$\textbf{i}_{0}$	& 
IF & 
	ID & EX & ME & WB\tikzmark{a} & {\large$\times$} 
	& {\large$\times$} & {\large$\times$} & {\large$\times$}\\
\cellcolor[gray]{0.8}$\textbf{i}_{1}$	& 
{\large$\times$} & 
	IF & ID & $\bigcirc$ & $\bigcirc$ & \tikzmark{b}EX & ME & WB\tikzmark{c} & 
	{\large$\times$}\\
$\textbf{i}_{2}$	& 
{\large$\times$} & 
	{\large$\times$} & {\normalsize$\square$} & {\normalsize$\square$} & 
	{\normalsize$\square$} &
	{\normalsize$\square$} & {\normalsize$\square$} &
	{\normalsize$\square$} & \tikzmark{d}IF \\
\bottomrule%
\multicolumn{10}{l}{}\\
\multicolumn{10}{l}{ME in SIMF instruction $\textbf{i}_1$ can take multiple cycles.}\\
\multicolumn{10}{l}{$\bigcirc$: insert pipeline stall.}\\
\multicolumn{10}{l}{{\normalsize$\Box$}: kill/delay the following instruction.}\\
\end{tabular}
\label{tab:pipeline}
\begin{tikzpicture}[overlay, remember picture, >=latex, 
yshift=.25\baselineskip, shorten >=.5pt, shorten <=.5pt]
    \draw [dashed,->] ({pic cs:a}) -- ({pic cs:b});
    \draw [dashed,->] ({pic cs:c}) -- ({pic cs:d});
\end{tikzpicture}
\end{table}%

Table~\ref{tab:pipeline} illustrates the pipeline control mechanism for SIMF with a classic pipeline time table. 
Column 1 shows the instruction sequence executed from top ($\textbf{i}_0$) to bottom. 
Row 1, shows the clock cycles starting from the left ($\textbf{CC}_0$) to right. In each clock cycle, at least one pipeline stage, i.e., IF, ID, EX, ME, and WB, is executed, unless the instruction is stalled (denoted as ``{\footnotesize$\bigcirc$}'') or delayed (``{\normalsize$\Box$}'').
As shown in Table~\ref{tab:pipeline}, for maintaining the program's correctness after flushing, SIMF ($\textbf{i}_1$) needs additional pipeline control mechanism to guarantee:
1) all the prior instructions must have been committed before flushing,
which is denoted as $\textbf{i}_0.\textrm{WB} \to \textbf{i}_1.\textrm{EX}$; and,
2) all the following instructions must observe the resultant state of the components in SoF,
which have been flushed, i.e., $\textbf{i}_1.\textrm{WB} \to \textbf{i}_2.\textrm{IF}$.

\subsection{SIMF Software}\label{sec_scheme}

SIMF can be deployed in software depending on the temporal isolation schemes. In this paper, we discuss two potential SIMF based temporal isolation schemes, which both use SIMF in kernel space (as a privileged instruction).

\textbf{Aggressive scheme:} Similar to~\cite{Zhang:2013:DRC}, where even the kernel space is not trusted, the SIMF instruction can be deployed at the boundary between the switch from user space to kernel space, and the switch from kernel space to user space. This way ensures that every kernel/user switch will incur flushing, even if there will be no process switch. Such a scheme creates a strict temporal isolation and can introduce high overhead due to much more frequent activation of flushing operations. 

\textbf{Moderate scheme:} Similar to~\cite{Ge:2019:TPM}, the SIMF instruction can be integrated in the software path where process/thread switch occurs, for instance, when the current security domain switches to another security domain in seL4 microkernel. This scheme will incur fewer flushing operations, while relying on the security assumptions provided by the security policy of the operating system.

Additional instructions are needed for RegFile flushing, to ensure the context will not be lost. If SIMF is deployed in the software path where context is saved, such as the trap entry of the operating system, the extra instructions needed will be much fewer. A few instructions are still required to be added to load back the values of the registers, which are needed for the system binary interface (SBI). Hence, for ease of deployment, register file flushing can be disabled.

\section{Evaluation}

In this section, we evaluate SIMF on a RISC-V processor, implemented on an FPGA with various workloads to answer the following questions.\\
How expensive the hardware cost is with SIMF implemented (\cref{sec_fpga})?\\
What is the performance penalty for the worst case, that is when L1 D-cache is full (\cref{sec_mmap})?\\
How SIMF performs  when inserted into a state-of-art microkernel (\cref{sec_sel4})?\\
How much does SIMF slow down  a contemporary Linux kernel with SMP in a multi-core scenario (\cref{sec_linux})?\\
How much does SIMF affect user programs (\cref{sec_mibench})?\\
What is the security enhancement (\cref{sec_sectest})?

\begin{table}[tb]\scriptsize
\centering
\caption{RISC-V System configuration}
\begin{tabular}{ll}
Attribute & Setting \\ 
\midrule
ISA			&	GC (IMAFDC) \\
L1D-Cache	&	32KiB 8-way 64B\\
L1I-Cache	&	32KiB 8-way 64B\\
L1D-TLB 		&	32 \\
L1I-TLB 		&	32 \\
L2-TLB 		&	128\\
BTB 		& 28	\\
BHT 		& 512\\
RAS 		& 6	\\
RAM		& 256MiB\\
\end{tabular} \label{tab_config}
\end{table}

To check the efficacy of SIMF, we implement the proposed SIMF as a new instruction, \textsc{Flushx}, in the Rocket core (i.e., in-order scalar RV64GC implementation of RISC-V ISA), which comes with Rocket Chip SoC (multi-core-capable and with hardware support for coherence)~\cite{Asanovic:EECS-2016-17}.

We use a Xilinx ZYNQ Ultrascale+ FPGA (ZCU102) board to implement the RISC-V processor(s), which technology-advanced and has greater amount of resources than the existing FPGA boards (e.g., ZC702, Zedboard, and ZYBO FPGAs) targeted by the original Rocket chip's FPGA build. We ported the original repository of Rocket chip's FPGA build to ZCU102. 
SIMF (\flushx{}) is implemented into the rocket core with the configuration specified in Table~\ref{tab_config}.

\subsection{Hardware Cost}\label{sec_fpga}

\begin{table}[!t]\scriptsize
\caption{FPGA utilization of \flushx{} and baseline system @\SI{180}{\mega\hertz}.
The overhead in brackets is without $O_{rf}$.}
\begin{center}
\begin{tabular}{c c c c c}
Resource & Baseline & \flushx{} & w/o $O_{rf}$& Overhead\\
\midrule%
LUT		& 37595 & 40107 & 38323 & 6.7\% (1.9\%) \\
F/F		& 16928 & 19436 & 17436 & 14.8\% (3.0\%) \\
BRAM		& 49 & 49 & 49 & -- \\
DSP		& 15 & 15 & 15 & -- \\
I/O		& 3 & 3 & 3& -- \\
BUFG		& 2 & 2 & 2& -- \\
MMCM		& 1 & 1 & 1& -- \\
\end{tabular}
\end{center}
\label{tab_resUtil}
\vspace{-10pt}
\end{table}%

The FPGA synthesis is performed with Vivado v2017.1 using default strategy.

Table~\ref{tab_resUtil} depicts the hardware cost, in terms of FPGA resources when the \flushx{} instruction is implemented. The major FPGA resources utilized include CLB LUTs, CLB flip/flops (F/Fs), block RAMs (BRAMs), DSPs (implementing arithmetic cells), physical input/output ports (IO), clock buffers (BUFG), and mixed mode clock manager (MMCM). This result shows that supporting \flushx{} instruction mainly increases the total LUTs (by 6.7\%) and F/Fs (by 14.8\%). 
We also observe that \flushx{} without $O_{rf}$ (equivalent to the complete functionality of existing temporal isolation) has minimal overheads (1.9\% LUTs and 3.0\% F/Fs).
The maximum clock frequency for \flushx{} is 187 MHz and for the baseline is 195 MHz.

\subsection{Flushing Overhead}\label{sec_mmap}

\begin{table}[!t]\scriptsize
\centering
\caption{Core-level flushing performance (ARMv8 implementation is hard IP of Cortex-A53 on ZCU102; RV64GC implementation is soft core on ZCU102; Intel x86 Haswell and ARMv7 Cortext-A9 in Yellow are obtained from~\cite{Ge:2019:TPM})}
\begin{tabular}{l c c c c}
Arch. & FO & \#Cycles & \#Dyn. Instr. & SoF \\ 
\midrule 
\rowcolor{Yellow}x86 & Indirect & 
	91800 & -- & L1-D\&I \\ 
\rowcolor{Yellow}ARMv7 & \texttt{dccisw}/\texttt{iciallu} 
	& 36000 & -- & L1-D\&I \\ 
ARMv8 & \texttt{dc\textvisiblespace cvau}/\texttt{ic\textvisiblespace ivau}&
	91299 & 63719 & L1-D\&I \\ 
RV64GC & \flushx{} & 16025.5 & 136 & Core-level \\ 
\end{tabular} \label{tab_mmap}
\vspace{-10pt}
\end{table}

The first case study aims to measure the performance overhead of the proposed core-level SIMF in comparison to the contemporary ISA support. The test program is manually designed to: 
one,  construct and fill a cache-sized contiguous memory space (via \smalltt{mmap()}); and,
two,  execute the core-level flush.

Table~\ref{tab_mmap} shows the comparison of overhead for core-level flushing between \flushx{} in the RISC-V processor and ARM Cortex-A53 using existing methods. 
ARM core flush is executed by calling \smalltt{\_\_clear\_cache()}, which targets the cache-sized memory space, provided by Linux to the user space, which is implemented in the same manner as the software functions in Figure~\ref{fig:arm_code}. \smalltt{\_\_clear\_cache()} essentially performs \smalltt{dc\textvisiblespace cvau}, i.e., data cache clean by virtual address (VA) to the point of unification (PoU), i.e., L2, and \smalltt{ic ivau} (instruction cache invalidate by VA to the PoU) instructions to flush L1 D-cache and I-cache. 

The results on the ARM are obtained by \smalltt{perf}. The test program is run 15,000 times to reduce the standard deviation, i.e., 593.5 (0.65\%) clock cycles and 249.9 (0.39\%) instructions. The result of RISC-V is obtained via inserted \smalltt{rdcycle} and \smalltt{rdinstret} instructions. The standard deviation for 100 runs of RISC-V is 82.1 clock cycles (0.51\%) and  0 instructions.

\flushx{} finishes flushing the complete core-level state with about 1/5 of the clock cycles used by ARMv8 (using user-mode flushing instructions), as well as about 1/2 and 1/5 clock cycles of ARMv7 and x86, reported in~\cite{Ge:2019:TPM}. \flushx{} reduces dynamic instruction count by a factor of 468, in comparison to ARMv8 (using user-mode flushing instructions).

\subsection{User Program Benchmark}\label{sec_mibench}

\begin{table}[t]\scriptsize
\caption{Geometric mean of performance overhead across MiBench}
\begin{center}
\begin{tabular}{c c c c}
System & $\Delta$CC & $\Delta$Dyn. Instr. & $\Delta$CPI\\
\midrule%
base-ecall		& \SI{0.58}{\percent} & \SI{-0.03}{\percent} & \SI{0.61}{\percent} \\
\flushx{}S		& \SI{0.76}{\percent} & \SI{0.002}{\percent} & \SI{0.76}{\percent} \\
\flushx{}T	& \SI{17.52}{\percent} & \SI{0.06}{\percent} & \SI{17.46}{\percent} \\ 
\end{tabular}
\end{center}
\label{tab_geomean}
\vspace{-10pt}
\end{table}%

\begin{figure}[!t]
\centering
\subfloat[Cycle count overhead]{
\includegraphics[width=\columnwidth]{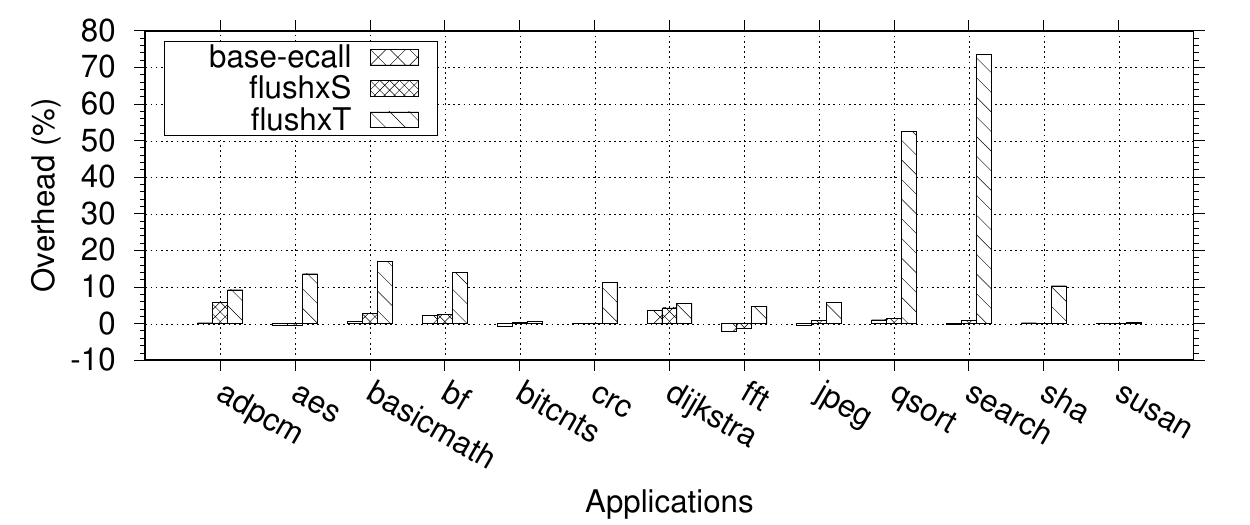}
\label{fig_medianRuntime}
}
\\
\vspace{-8pt}
\subfloat[Dynamic instruction count overhead]{
\includegraphics[width=\columnwidth]{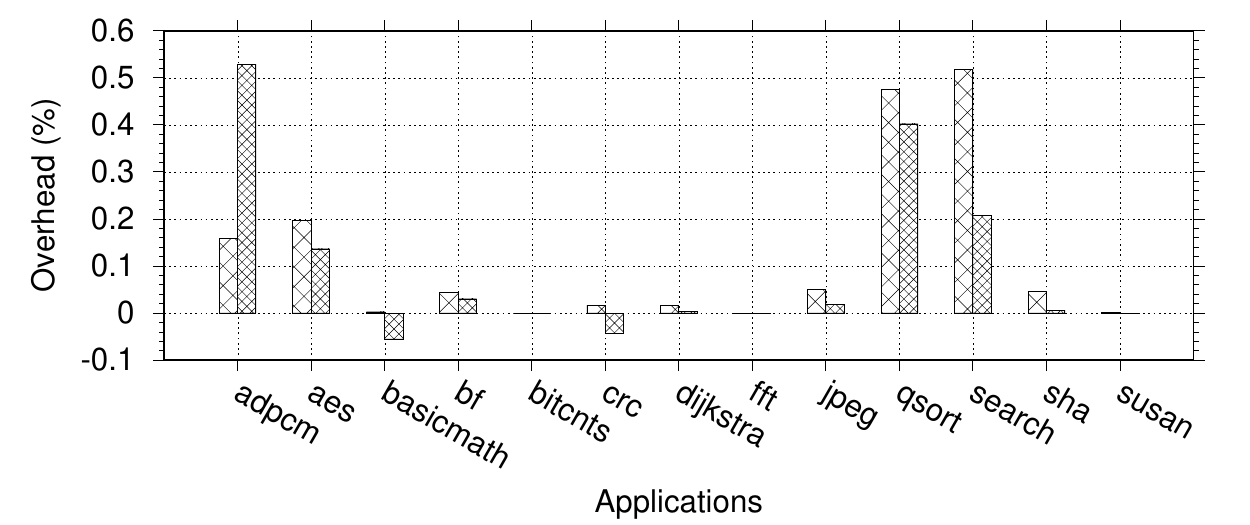}
\label{fig_medianInstr}
}
\caption{Median performance overhead (in percentage) of running MiBench.} 
\label{fig_mibench}
\vspace{-10pt}
\end{figure}

To observe the runtime overhead of executing \flushx{} instruction in real-world user programs, we test our systems using a representative embedded application benchmark suite, MiBench~\cite{Guthaus:2001:MFC}. We test \flushx{} with two scenarios:
1) \flushx{}S incur a \flushx{} via a syscall (similar to moderate scheme);
and, 2) \flushx{}T incur \flushx{} at every kernel/user crossing (similar to aggressive scheme).
The benchmark suite is run on a simple application execution environment (riscv-pk), which provides user programs with POSIX syscall services. riscv-pk is customized to deploy \flushx{} in a new syscall and trap entry. For each program, we execute the program 50 times (standard deviation $\leq$ 1\%).

As shown in Table~\ref{tab_geomean}, performing \flushx{} instruction once per execution (\flushx{}S) incurs negligible overhead in both clock cycles and dynamic instruction count (within the margin of error), especially when compared to pure syscall overhead (``base-ecall'').
Performing \flushx{} aggressively at every entering kernel space (\flushx{}T) can substantially increase the execution time and CPI by 15\%. The main reason for the difference of overhead between \flushx{}S and \flushx{}T is that \flushx{}T scheme activates a greater number of \flushx{} execution in these programs from the original system calls. These system calls do not include \flushx{} in the execution path at runtime in \flushx{}S system.

Figure~\ref{fig_medianRuntime} and \ref{fig_medianInstr}
depict the detailed view of the overhead for each benchmark application.
For cycle overhead (shown in Figure~\ref{fig_medianRuntime}), among all the benchmark programs, 
there are two prominent cases, where \flushx{}T incurs a substantial cycle overhead. These are  in stringsearch \smalltt{search} (around 70\%) and quick sort \smalltt{qsort} (roughly 50\%). The reason is that these two programs proportionally have more syscalls.
In some cases, e.g., bitcounts \smalltt{bitcnts}, FFT \smalltt{fft}, and JPEG \smalltt{jpeg}, the cycle overhead is negative, because the overhead is negligible within the range of standard deviation (due to cache and branch prediction behaviors).
For dynamic instruction count, the overhead is negligible across all the benchmark programs.

\subsection{seL4 Microkernel}\label{sec_sel4}


\makeatletter
\newcommand{\mypm}{\mathbin{\mathpalette\@mypm\relax}}
\newcommand{\@mypm}[2]{\ooalign{%
  \raisebox{.1\height}{$#1+$}\cr
  \smash{\raisebox{-.6\height}{$#1-$}}\cr}}
\makeatother

\begin{table}[!t]\scriptsize
\centering
\caption{seL4 kernel scheduling overhead. RISC-V runs seL4 v10.1.1 kernel with seL4test workloads. In scheduling function, four execution paths incurring thread switch are measured respectively. ARM data in yellow is obtained from \cite{Ge:2019:TPM} measuring domain switch time.}
\begin{tabular}{l c c<{\hspace{-\tabcolsep}}>{\hspace{-\tabcolsep}\,}c
<{\hspace{-\tabcolsep}\,}>{\hspace{-\tabcolsep}}c  c@{\hskip3pt}c@{\hskip3pt}c}
 & & \multicolumn{3}{c}{\#Cycles} & \multicolumn{3}{c}{\#Dyn. Instr.}\\
Arch. & FO & mean & $\mypm$ & std. & mean & $\mypm$ & std.  \\ 
\midrule
\rowcolor{Yellow}
ARMv7 & -- & $560$ & $\mypm$ & $1\%$ & \multicolumn{3}{c}{--} \\
\rowcolor{Yellow}
ARMv7 & Ge et al.\cite{Ge:2019:TPM} & $21600$ & $\mypm$ & $1\%$ & \multicolumn{3}{c}{--} \\
RV64GC & -- & $940.5$ & $\mypm$ & $318.7$ & $432.8$ & $\mypm$ & $155.3$ \\ 
RV64GC & \flushx{} & $8568.1$ & $\mypm$ & $3738.5$ & $436.9$ & $\mypm$ & $147.1$\\
\end{tabular} \label{tab_sched}
\vspace{-10pt}
\end{table}

\begin{figure}[tb]
\centering
  \subfloat[Original]{
  \includegraphics[width=0.5\columnwidth]{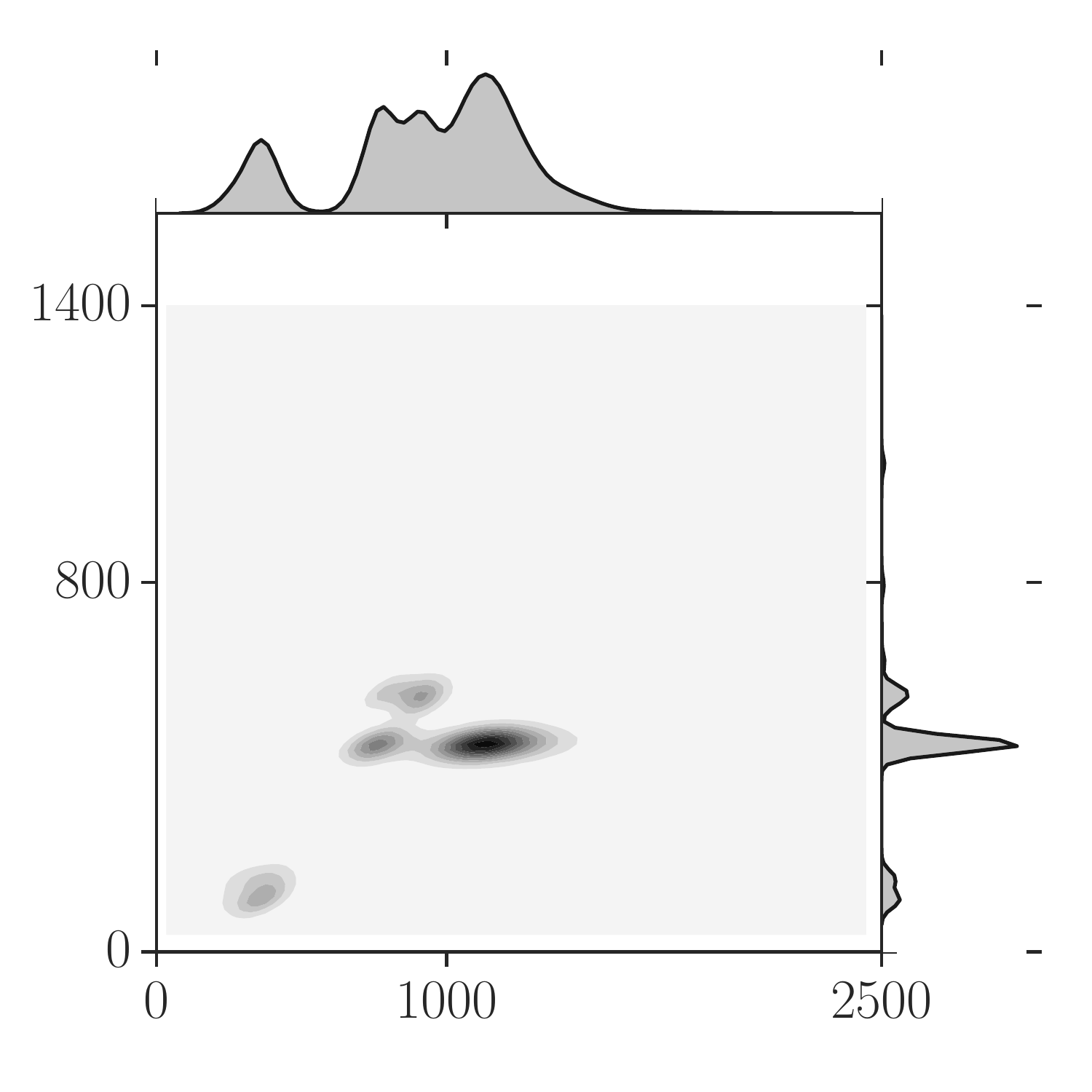}
  \label{fig:origSched}
  }
    \subfloat[\flushx{}]{
  \includegraphics[width=0.5\columnwidth]{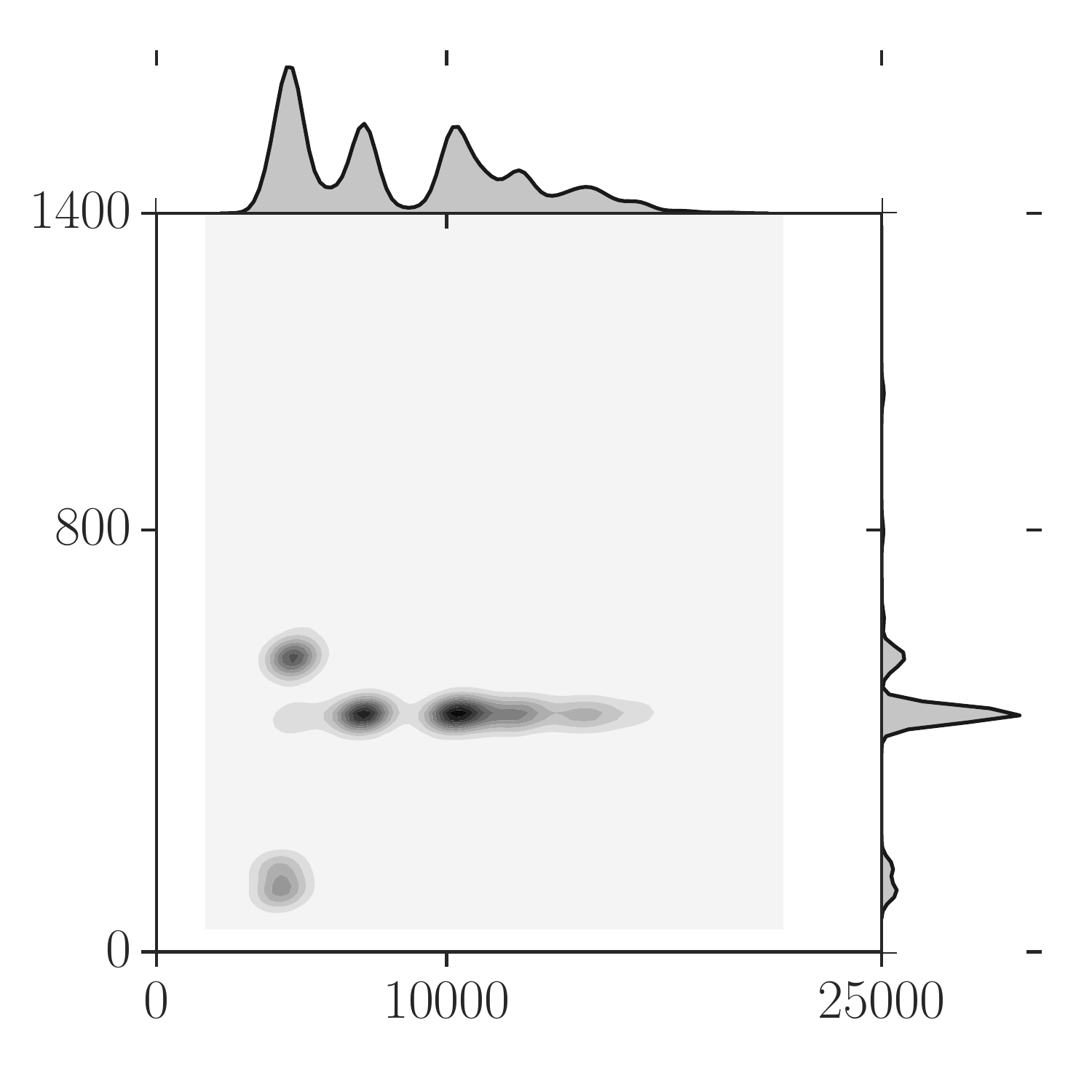}
  \label{fig:flushxSched}
  }
  \caption{Distribution of seL4 kernel scheduling cost
   with thread switch incurred, w/ and w/o \flushx{}. X-axis is cycle count. 
   Y-axis is dynamic instruction count. 
   The colored area in X-Y plane denotes the two-dimensional density function
   from kernel density estimation (KDE). The shaded curves in the marginal area at top
   and right denote one-dimensional density function with respect to X and Y.} 
\label{fig:sched}
\vspace{-12pt}
\end{figure}

To observe the system-level manifestation of SIMF in a real-world microkernel.
We added \flushx{} into seL4 microkernel (\texttt{v10.1.1}), which is a security-oriented capability-based microkernel. 
We insert \flushx{} instruction into functions for thread switch, such as \smalltt{switchToThread()} and\\ 
\smalltt{switchToIdleThread()}.
seL4's domain switch is a special case of thread switch, which largely follows the same execution path as thread switch.

To quantify the performance of SIMF, we measured the cost of scheduling, where thread switch is invoked, with the seL4test as the workload. seL4test is composed of 100 testing programs, each of which targets one set of seL4 properties.
During the execution of seL4test, \smalltt{schedule()} is called 3,496,281 times, while 427,189 events of thread switch are incurred during schedule. Both systems with and without \flushx{} finish the seL4test correctly, with the same output score.

Table~\ref{tab_sched} shows the total clock cycles and dynamic instruction count of scheduling for the system with and without \flushx{} instruction implemented, in comparison to the domain switch cost reported in~\cite{Ge:2019:TPM}, where ARMv7 's flushing mechanism is adopted. The cycle count of ARMv7 is calculated based on time and clock speed reported.
On average, ARMv7's ISA support results in about 38.57x overhead (in comparison to the baseline) in clock cycles, while \flushx{} incurs less than 10x cycle overhead with 4 additional instructions. As shown in Figure~\ref{fig:sched}, the cycle cost varies depending on the actual execution path in scheduling (some paths are affected more by \flushx{}). The distribution of the dynamic instruction count is minimally affected by \flushx{}.

\subsection{Linux with lmbench}\label{sec_linux}

\begin{figure*}[tb]
\centering
  \subfloat[32KiB w/ L2TLB, \#Core=1, GM=2.57]{
  \includegraphics[width=0.33\textwidth]{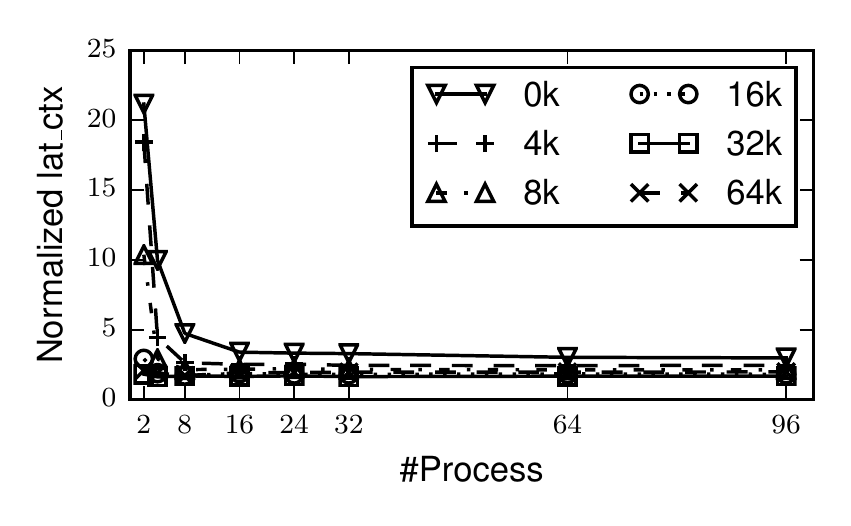}
  \label{fig:ctx1p}
  }
  \subfloat[32KiB w/ L2TLB, \#Core=4, GM=2.70]{
  \includegraphics[width=0.33\textwidth]{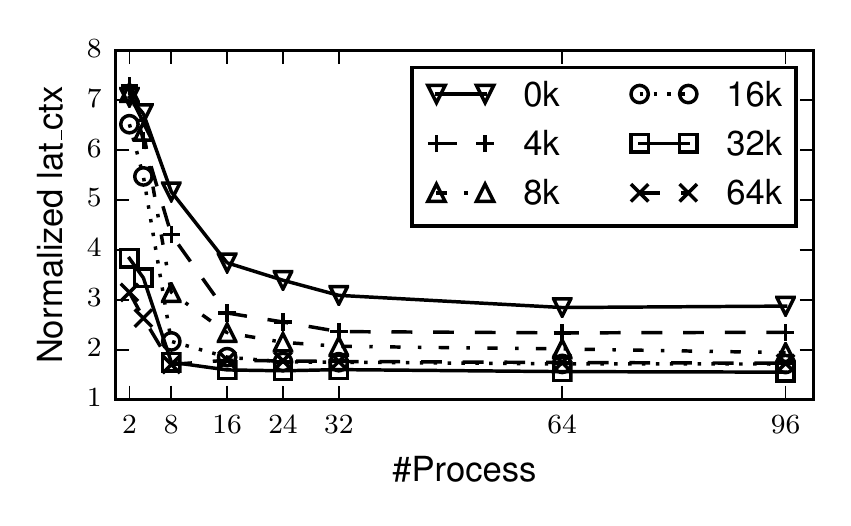}
  \label{fig:ctx4p}
  }
  \subfloat[32KiB w/ L2TLB, \#Core=8, GM=3.09]{
  \includegraphics[width=0.33\textwidth]{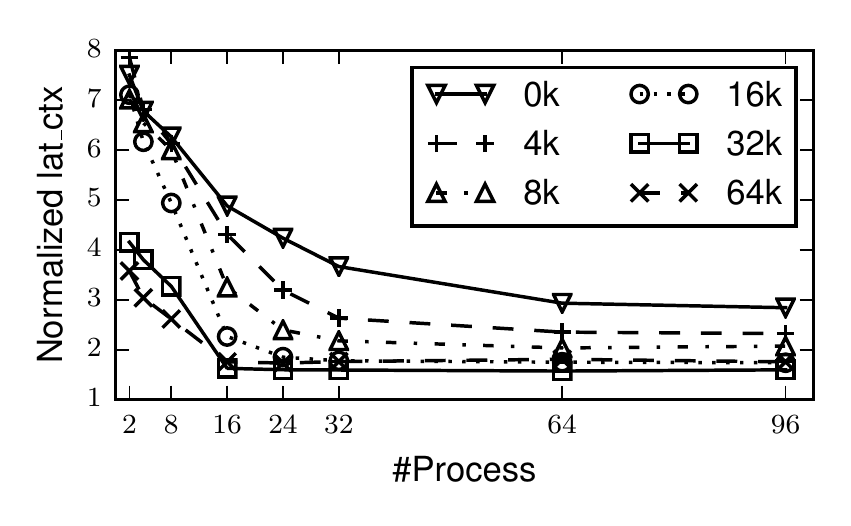}
  \label{fig:ctx8p}
  }
  \\
  \subfloat[16KiB w/o L2TLB, \#Core=1, GM=1.65]{
  \includegraphics[width=0.33\textwidth]{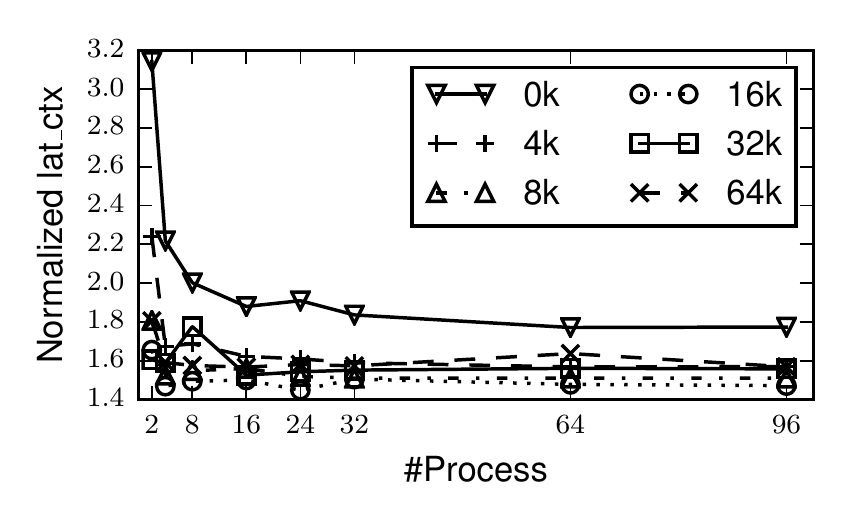}
  \label{fig:ctx1p}
  }
  \subfloat[16KiB w/o L2TLB, \#Core=4, GM=1.60]{
  \includegraphics[width=0.33\textwidth]{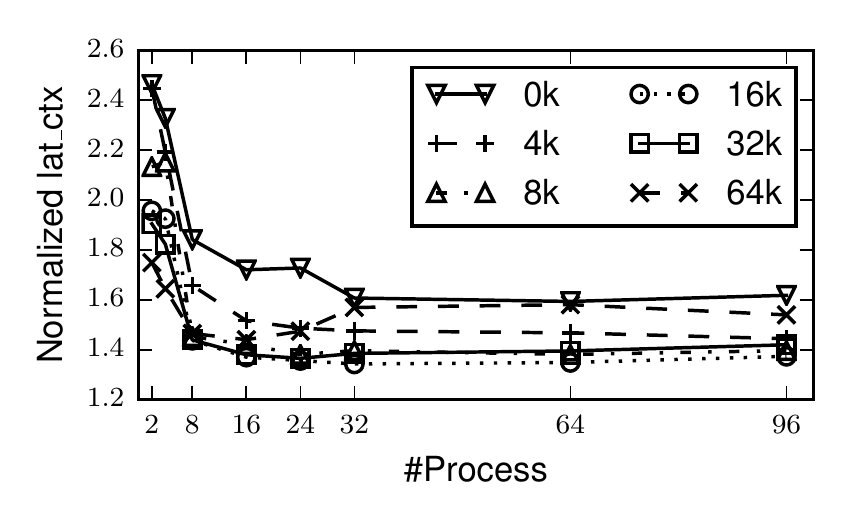}
  \label{fig:ctx4p}
  }
  \subfloat[16KiB w/o L2TLB, \#Core=8, GM=1.62]{
  \includegraphics[width=0.33\textwidth]{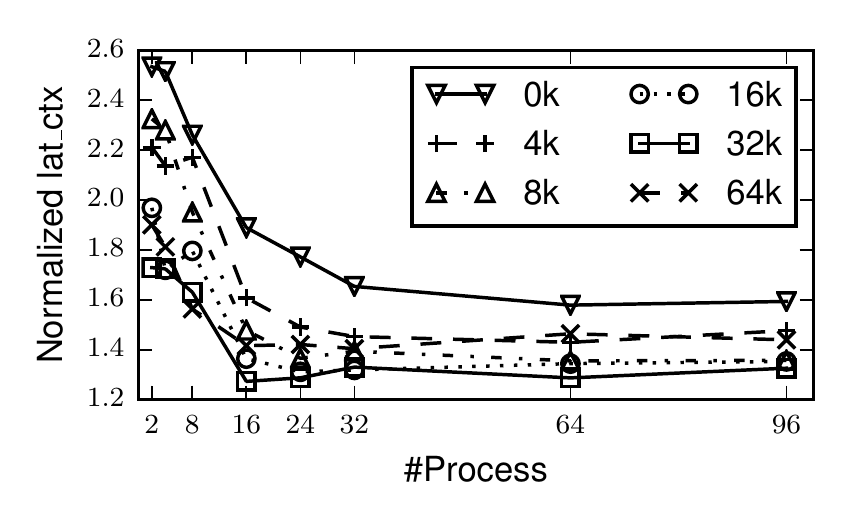}
  \label{fig:ctx8p}
  }
  \caption{lmbench contex switch latency (lat\_ctx) as a function of the number of processes (2, 4, 8, 16, 32, 64, 96), with different size of working set, normalized to the baseline. GM: geometric mean. RISC-V Linux kernel version is 4.20 with \flushx{} inserted in \texttt{\_\_switch\_to()}. Each line represents a working set size in the legend.} 
\label{fig:lmbench}
\end{figure*}

We choose RISC-V Linux kernel v4.20 and add \flushx{} into the thread switch procedure \smalltt{\_\_switch\_to}. We test the \flushx{} Linux and compare with the baseline Linux kernel by running lmbench~\cite{McVoy:1996:LPT}. For both systems, we create three types of hardware, composing 1 core, 4 cores, and 8 cores, each with two configurations of caches (16 KiB and 32 KiB) and TLBs (with and without L2 TLB).

Figure~\ref{fig:lmbench} shows the overhead of context switch latency due to \flushx{}. For larger caches and TLBs, \flushx{} incurs more overhead with a smaller working set and fewer processes. Using smaller caches and TLBs leads to much smaller flushing overhead (almost 1/2 in 8-core system). It is also seen that as processes increase, the overhead will approach 50\% (for large caches and TLBs) or less (for small caches and TLBs). The worst-case overhead is witnessed in an extreme scenario, where two processes runs on one core with larger caches and TLBs, computing with a very small working set (0KB).

\subsection{Temporal Isolation Test}\label{sec_sectest}

\begin{figure}[!t]
\vspace{-10pt}
\centering
\subfloat[Baseline]{
\includegraphics[width=.9\columnwidth]{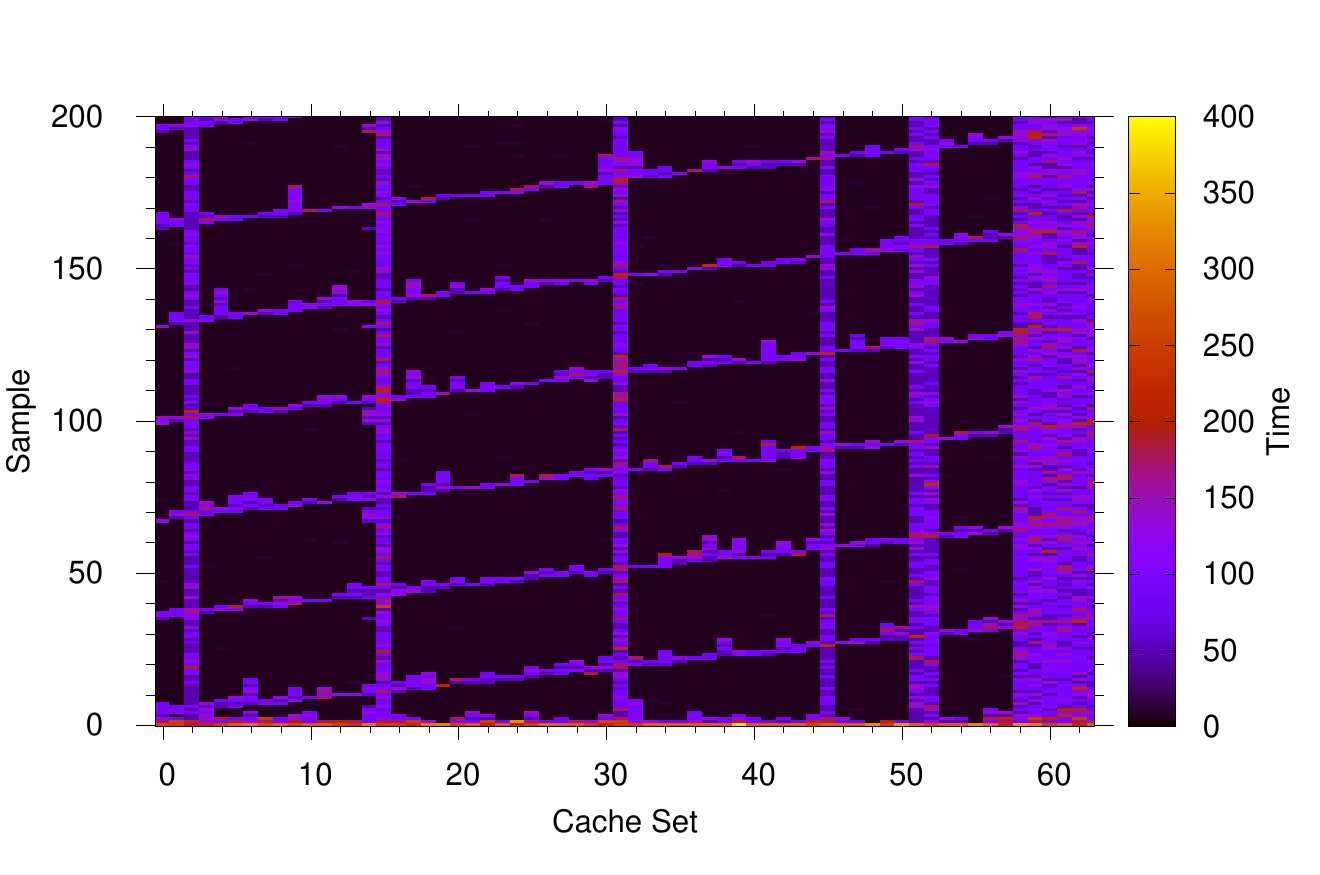}
\label{fig_mapbs}
}
\\
\vspace{-18pt}
\subfloat[\flushx{}]{
\includegraphics[width=.9\columnwidth]{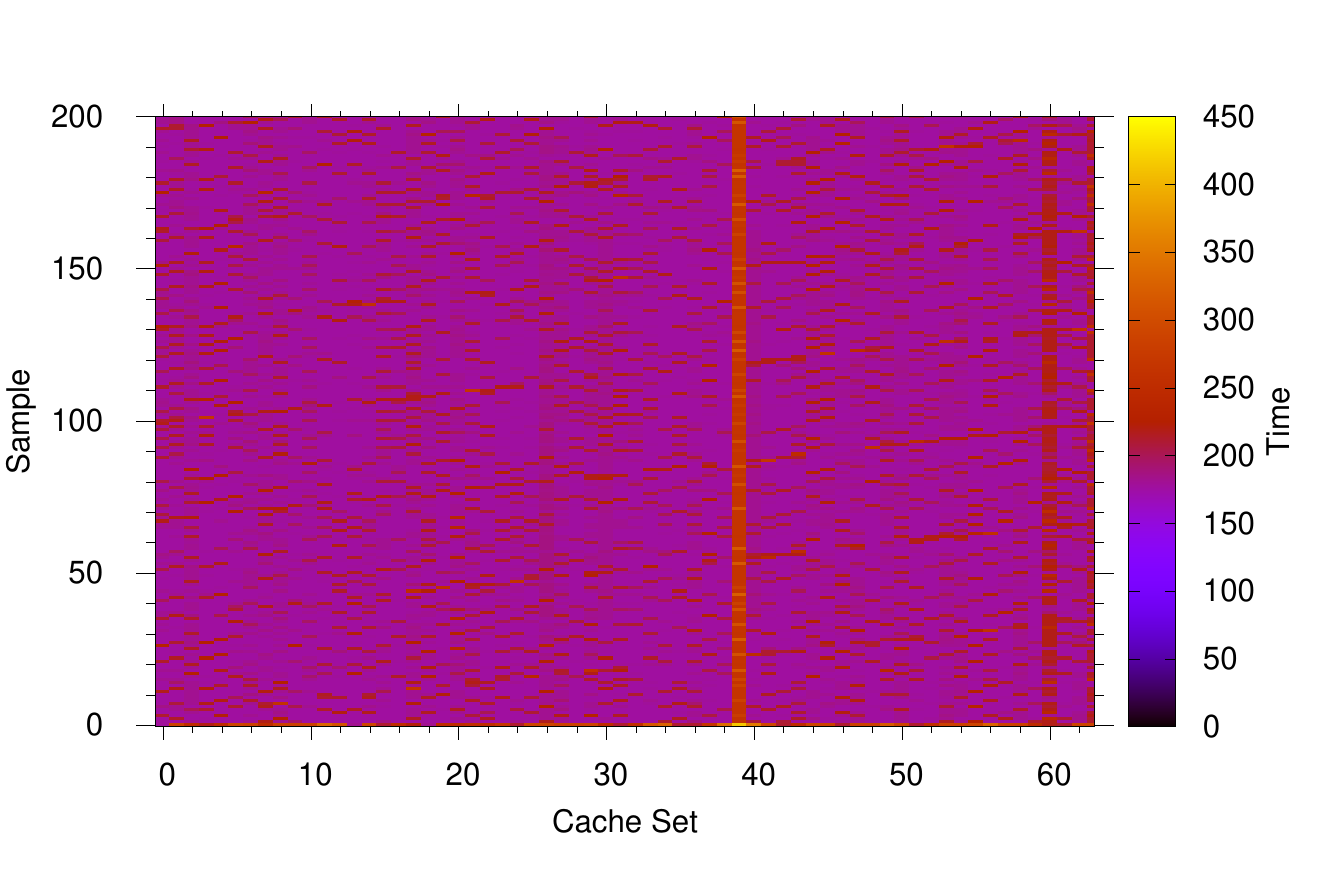}
\label{fig_mapfx}
}

\vspace{-4pt}
\caption{Timing channel of L1 D-cache by \smalltt{Prime+Probe}.} 
\label{fig_attack}
\vspace{-10pt}
\end{figure}

To evaluate the effects of \flushx{} instruction in face of timing channel attacks, we implement a classic L1 D-cache attack, namely \smalltt{Prime+Probe} based on the open-source tool called Mastik~\cite{Yarom:mastik}. We port Mastik to RISC-V to mount \smalltt{Prime+Probe} cache timing attacks.

We create a simplified attack using lightweight proxy kernel, PK (riscv-pk).
We implement the victim program as a dedicated system call. The content of the victim program is a patterned data cache access, adapted from the example code from Mastik. \flushx{} is inserted before returning to attacker' process on \flushx{} core.
We modify the attack program to raise this special system call, after cache prime is finished.
The probed samples are set to 20,000, which is sufficient to witness the significance of the cache channels.

Figure~\ref{fig_mapbs} and \ref{fig_mapfx} illustrates the samples from Sample 1 to 200, in the cache timing attack, on the baseline and \flushx{} systems, respectively. The clear cache behavior patterns of cache set accesses can be observed for the unprotected baseline system, where the cache set hit (less than 10 cycles of cache access time, in black) and miss (more than 50 clock cycles of access time in purple) can be differentiated.
Figure~\ref{fig_mapfx} shows that the protected system, which executes \flushx{} before returning to the attacker's process, and can cleanse the residual state (all cache misses) in the L1 D-cache.

\section{Related Work}

Timing attack countermeasures have been proposed at differing layers of the computing stack, from hardware right up to the application layer.

\textbf{Hardware-based} countermeasures~\cite{Werner2019,Wang2008,Page2005,Qureshi2018,Trilla2018} mostly focus on redesigning the cache  against cache based timing channel attacks.
The work in~\cite{Page2005} enforces cache partitioning to remove cache contention.
Newcache~\cite{Wang2008} introduces randomized mapping in replacement of cache partitioning for better flexibility.
CEASE/CEASER~\cite{Qureshi2018} uses encrypted cache addressing scheme to implement efficient randomization.
\cite{Trilla2018} and ScatterCache~\cite{Werner2019} extends encryption-based randomization to combine cache address and process ID as encryption input. 
Aiming at speculation-based side channel attacks, CleanupSpec~\cite{Saileshwar2019} modifies cache policies and coherency management on system bus, in order to roll back the cache state after miss-speculation.
These methods extensively modify the original cache operations, largely target cache timing attacks at the last level cache (LLC), and are evaluated in abstract models using system simulators (e.g., GEM5~\cite{Binkert2011}). No hardware implementation has been reported.
Another line of hardware-based countermeasures~\cite{Li:2014:SLH,Ferra2018} seek to eliminate timing channels by enforcing information flow tracking in hardware. These methods introduce new hardware description languages and usually incur additional memory elements for security tags.

\textbf{Software-based} methods are usually implemented in hypervisors, operating systems, or applications. These methods can categorized as follows:
Based on time measurement, FuzzyTime~\cite{Hu1991a} and TimeWarp~\cite{Martin2012} adds intentional noise to the system clock so as to prevent the adversary from accurately measure the timing of microarchitectural events. The time noise will also affect the original programs system-wide, wherever accurate time is required.
StealthMem~\cite{Kim2012,Dong2018} implements spatial partitioning to enforce information isolation in LLC against timing attacks.

\textbf{Flushing} mechanism has been adopted by a variety of OS-level software-based countermeasures~\cite{Zhang:2013:DRC,Ge:2019:TPM}. 
D\"uppel~\cite{Zhang:2013:DRC} uses flushing of private L1 and L2 caches to inject timing noise so as to maximize the difficulty of mounting cache-based timing attacks. D\"uppel assumes the attackers can perform a probe as frequently as every 50,000 (about 60 KHz at 3 GHz clock speed) to 90,000 CPU cycles.
Ge et al~\cite{Ge:2019:TPM} leverages core-level flushing along with L2 and LLC cache partitioning to minimize the available timing channels in all the microarchitectural components. The flushing is assumed to be performed at every domain switch.

Varys~\cite{Oleksenko:2018:VPS} uses flushing for cleansing private L1 and L2 caches, to protect Intel SGX enclave from various timing channel attacks, assuming the extreme case that privileged software is controlled by the attacker. The flushing is performed at the frequency of 100 Hz, 5.5 KHz, and 10 KHz. For 100 Hz, Varys incurs 19\% drop-down in throughput of Nginx due to indirect flushing in Intel ISA.

MI6~\cite{bourgeat2019} is an enclave design for out-of-order RISC-V processors, which also includes a \texttt{purge} instruction to flushing on-core state (excluding register-file) upon (de)scheduling. 
SIMF differs to MI6 in several aspects: MI6 is specific to only RiscyOO processor and cannot be generalized, while SIMF can be implemented for any in-order processor; MI6 relies on the specific features of the baseline processor RiscyOO and does not have the mechanism to actually clean  replacement tags in caches and TLBs; MI6 does not actually clean some microarchitectural states (e.g., issue queue), which may well be exploited by novel attacks; and, MI6's flushing mechanism does not write back the dirty cache lines, hence, it does not work for  caches with write-back policy (which is dominant in real-world modern processors).

\section{Conclusion}
In this paper, we have presented SIMF, a new ISA extension to support efficient temporal isolation. SIMF is capable of flushing core-level state in one instruction execution and can be integrated with OS-level timing attack mitigation. We have prototyped SIMF as a \flushx{} instruction on RISC-V processor and evaluated on FPGA with a real-world microkernel, Linux kernel, user programs, and \smalltt{Prime+Probe} cache timing attack. Our evaluation shows that SIMF removes the timing channels effectively with significantly less clock cycles and dynamic instruction count.

\bibliographystyle{IEEEtranS}
\bibliography{main}

\end{document}